\documentclass{PoS}

\newcommand{\be}{\begin{equation}}
\newcommand{\ee}{\end{equation}}

\title{Computation of the string tension in three dimensions using large N reduction }

\ShortTitle{Computation of the string tension in three dimensions using large N reduction }

\author{\speaker{Joseph Kiskis}\\
    Department of Physics, University of California,
    Davis, CA 95616, USA \\
        E-mail: \email{jekiskis@ucdavis.edu}}

\author{Rajamani Narayanan\\
  Department of Physics, Florida International University,
Miami, FL 33199, USA\\
        E-mail: \email{rajamani.narayanan@fiu.edu}}

\abstract{We numerically compute the string tension in the large N 
limit of three dimensional Yang-Mills theory using Wilson loops. 
Space-time Wilson loops are formed using smeared space-like links and 
unsmeared time-like links. We use partial reduction and both unfolded 
and folded Wilson loops in the analysis.}

\FullConference{The XXVI International Symposium on Lattice Field Theory \\
		 July 14 - 19, 2008\\
		 Williamsburg, Virginia, USA}

\begin{document}

\section{Introduction}

The method of large $N$ continuum 
reduction~\cite{Narayanan:2003fc,Kiskis:2003rd} for $SU(N)$ gauge theory 
allows for the calculation of the infinite volume, infinite $N$ limit of 
certain physical quantities using volumes reduced to a small physical 
size. Numerical estimates~\cite{Narayanan:2003fc,Kiskis:2003rd} of the 
physical critical size above which continuum reduction holds indicate 
that this method can be used to produce practical results. The chiral 
condensate~\cite{Narayanan:2004cp} and the pion decay 
constant\cite{Narayanan:2005gh} were calculated in the large $N$ limit 
in four dimensions using continuum reduction. In this paper, we show 
that the method can be extended beyond bulk quantities and that it also 
produces reliable results for quantities with space-time dependence such 
as the heavy quark potential, from which the string tension can be 
extracted. Invoking large $N$ continuum reduction, we included Wilson 
loops larger than the size of the lattice. The results validate the 
method of continuum reduction for calculating quantities based on the 
space-time dependence Wilson loops.

A precise calculation of the string tension in three dimensional $SU(N)$ 
gauge theories has been performed with $N$ up to 8 on large 
lattices~\cite{Bringoltz:2006zg}. We present a complementary calculation 
with $N=47$ on $5^3$ lattices using continuum reduction. The calculation 
of Ref.~\cite{Bringoltz:2006zg} used correlation functions of smeared 
Polyakov loops to extract the string tension. After extrapolating to 
$N=\infty$ and to the continuum, the result was
\be
\frac{\sqrt\sigma}{g^2N} = 0.1975 \pm 0.0002 - 0.0005
\label{bt1}
\ee
where $g$ is the gauge coupling. This has to be compared with the 
analytical calculation in~\cite{Karabali:1998yq}, namely, 
$\frac{1}{\sqrt{8\pi}}\approx0.1995$. Although the two results are not 
in perfect agreement, the main observation is that the approximations 
used in the analytical calculation are very well motivated.

Our use of continuum reduction to directly compute the $N=\infty$ limit 
of the string tension by working at large enough $N$ so that the finite 
$N$ corrections are smaller than the numerical errors 
gives~\cite{Kiskis:2008}
\be
\frac{\sqrt\sigma}{g^2N} = 0.1964 \pm 0.0009 .
\label{thisresult}
\ee
This result and that of (\ref{bt1}) are consistent at the level of their 
one sigma errors. This level of agreement is, in turn, consistent with 
neither the large $N$ extrapolation of Ref.~\cite{Bringoltz:2006zg} nor 
the volume reduction of the present calculation having unexpected 
errors. While both of the numerical results lie below the analytical 
estimate, the discrepancy is relatively small. Thus the numerical 
evidence that the analytical result is an excellent first approximation 
that captures much of the physics remains strong.

The paper is organized as follows. We explain how we use smeared Wilson 
loops to compute the string tension in Section~\ref{details}. The 
lattice results for the string tension along with the continuum 
extrapolation are also presented in this section. An intermediate step 
in our calculation is the dimensionless ground state string energy 
$m(k)$. In Section~\ref{string}, we show results for $m(k)$ at one fixed 
lattice coupling to illustrate its behavior as a function of $k$ and how 
it is used to extract the string tension. We also show that $m(k)$ is 
unaffected by the smearing parameter. We illustrate the extraction of 
$m(k)$ at one fixed coupling in Section~\ref{mass}. Here we show how the 
smearing parameter affects the overlap with the ground state. The main 
result in this paper is obtained using $N=47$. We show that the finite 
$N$ and finite volume corrections are small at this value of $N$ in 
Section~\ref{finiteN}. We explain why this method is preferred over the 
Creutz ratio in Section~\ref{crratio}.

\section{String tension using Wilson loops and continuum
reduction}\label{details}

\begin{figure} 
\centering 
%%----start of first figure---- 
\begin{minipage}[t]{0.4\linewidth} 
\centering 
\includegraphics[width=2.4in,clip=true]{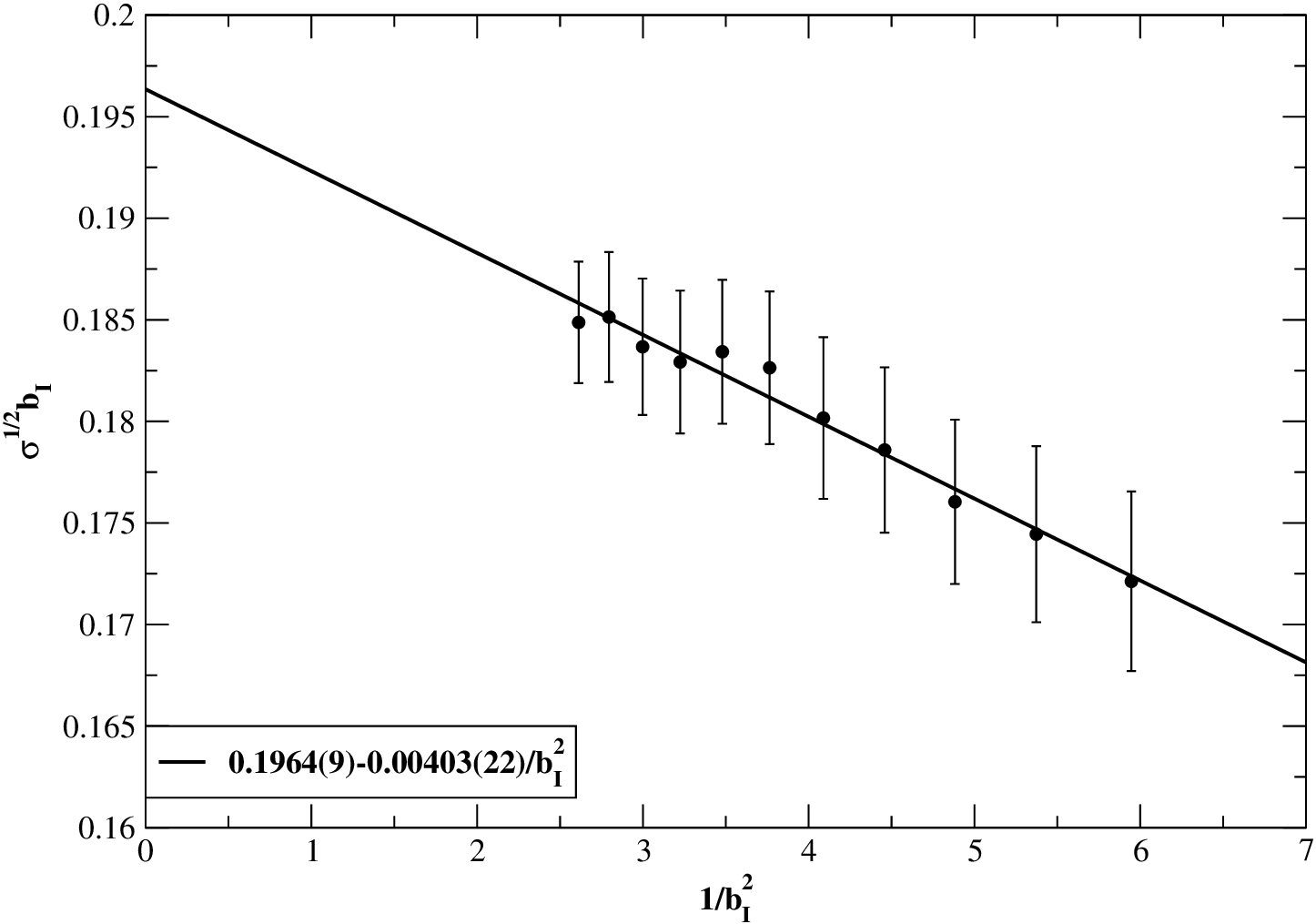} 
\caption{
The string tension is plotted as a function of the
lattice spacing $b_I^{-1}$. The fit is an extrapolation to
the continuum. 
}
\label{stringcont} 
\end{minipage}% 
\hspace{1cm}% 
%%----start of second figure---- 
\begin{minipage}[t]{0.4\linewidth} 
\centering 
\includegraphics[width=2.4in,clip=true]{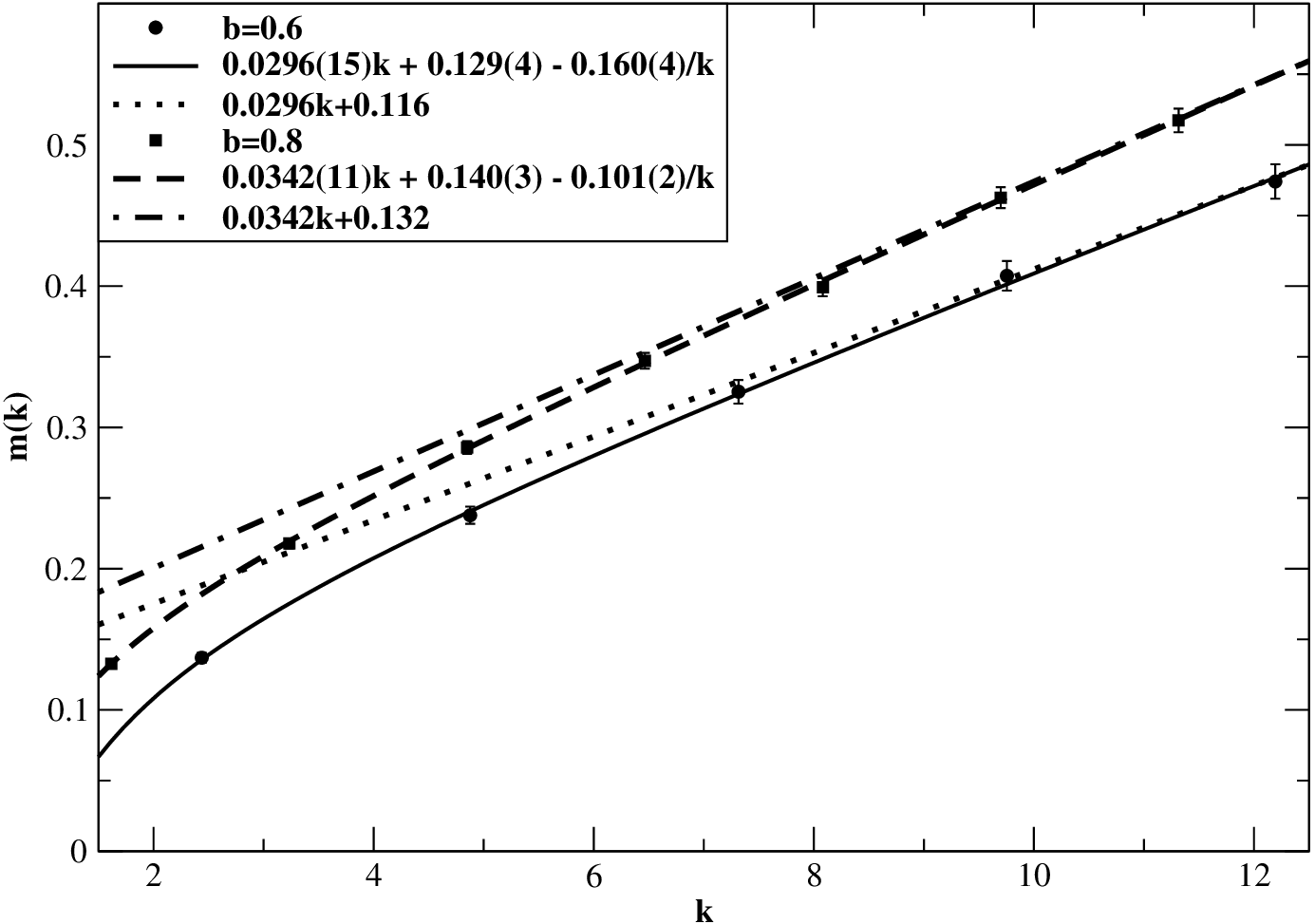} 
\caption{
The ground state energy $m(k)$ as a 
function of $k$ for the coarse and fine lattice
spacings considered here.
} 
\label{tension} 
\end{minipage} 
\end{figure}

Consider $SU(N)$ Yang-Mills theory on a periodic lattice with the 
standard Wilson gauge action. The method of~\cite{Bringoltz:2006zg} is 
to measure the string tension using correlations of Polyakov loops with 
separation $t$ that wind around a space direction. Continuum 
reduction~\cite{Narayanan:2003fc,Kiskis:2003rd} implies that the large 
$N$ Yang-Mills theory in a continuum box of size $l^3$ is independent of 
$l$ as long as $ l > l_c = 1/T_c$ with $T_c$ being the deconfining 
temperature. One should be able to compute expectation values of Wilson 
loops of arbitrary size on an $l^3$ continuum box using folded Wilson 
loops and extract the string tension. To implement this approach to the 
three-dimensional Yang-Mills theory string tension, we use the following 
procedure:
\begin{itemize}
\item We fix the lattice size to $L^3$. We use $L=5$ for
the most part and only use $L=4$ to verify reduction.
\item We fix $N$ so that finite $N$ corrections are small.
We set $N=47$ and show using one instance that
finite $N$ corrections are small at $N=47$.
\item We pick an appropriate range of lattice coupling
$b=\frac{1}{g^2N}$.
\begin{itemize}
\item $b$ cannot be too small since we have to be away
from the bulk transition on the lattice associated
with the development of gap in the eigenvalue distribution
of the plaquette operator~\cite{Bursa:2005tk}. Therefore,
we pick $b\ge 0.6$.
\item $b$ cannot be too big since we have to be below
the deconfining transition for $L=5$. Therefore, we pick
$b \le 0.8$~\cite{Narayanan:2007ug}.
\end{itemize}
\item We use smeared space-like links and unsmeared
time-like links.
\item We use the tadpole improved coupling $b_I=be(b)$
to set the scale
and consider $K \times T$ Wilson loops $W(K,T)$ with 
$1.5 < \frac{K}{b_I},\frac{T}{b_I} < 12.5$.
This amounts to expectation values of Wilson loops that range from
$0.82$ to $2\cdot 10^{-4}$.
\item Keeping $K$ fixed, we fit 
\be
ln W(k,t) = - a - m(k) t;
\ee
where
$ k=\frac{K}{b_I}$ and $t=\frac{T}{b_I}$ are the dimensionless
extent in the space and time direction respectively.
$m(k)$ is the dimensionless ground state energy.
This fit assumes that there is a perfect overlap with the
ground state. Note that 
$a$ should be zero since $W(k,0)=1$. Any small deviation from
zero seen in the fit is due to the contribution from
excited states. 
\item Finally, $m(k)$ is fit to
 $\sigma b_I^2 k + c_0b_I + \frac{c_1}{k}$.
The combination $\sqrt{\sigma}b_I$ is plotted as a function
of $b_I^{-2}$. We expect lattice spacing
effects to lead off as $b_I^{-2}$ in Yang-Mills theories and
this is indeed the case in Fig.~\ref{stringcont}. The
continuum limit extracted from this figure was quoted 
in~Eqn.(\ref{thisresult}).
\end{itemize}

The use of smeared links improves the
measurement of Wilson loops. 
They enhance the overlap of the space-like sides of the Wilson loops 
with the ground state. This increases the signal relative to the 
fluctuations and simplifies the $t$ 
behavior of the loops~\cite{Teper:1998te}. One step in the iteration
takes one from a set $U^{(i)}_k (x_1,x_2,t)$ to a set 
$U^{(i+1)}_k (x_1,x_2,t)$. Before reunitarization, the weight 
of $U^{(i)}_k (x_1,x_2,t)$ 
is $(1-f)$ while that of each staple is $f/2$.
The time-like links, $U_3(x_1,x_2,t)$, are not smeared,
and the smearing only involves space-like staples.
There are two parameters, namely, the smearing factor $f$
and the number of smearing steps $n$. Only the product
$\tau=fn$ matters, and $f$ plays the role of a discrete smearing
step. For a given $\tau$, the overlap of the
smeared loop with the ground state does not depend on $f$
as long as it is small. But the overlap of the
smeared loop with the ground state does depend upon $\tau$.
We set the value of the smearing parameter
to $\tau=2.5$ by choosing $f=0.1$ and $n=25$.
To study the effect of varying $\tau$, we also consider
$\tau=1.25$ ($f=0.05$ and $n=25$) at one coupling.

\section{Extraction of string tension}\label{string}

$SU(N)$ gauge fields were generated on a $5^3$ periodic lattice
using the standard Wilson action. One gauge 
field update of the whole lattice~\cite{Kiskis:2003rd} is
one Cabibbo-Marinari heat-bath update of the whole lattice
followed
by one $SU(N)$ over-relaxation update of the whole lattice.
A total of $1500$ such updates were used to achieve
thermalization. Measurements were separated by $10$ such
updates and all estimates are from a total of $832$ such
measurements. Errors in all quantities at a fixed $b$
and $N$ were obtained by jackknife with
single elimination.

The ground state energy $m(k)$ obtained as a function
of $k=\frac{K}{b_I}$ is fit to
\be
m(k) = \sigma b_I^2 k + c_0 b_I + \frac{c_1}{k}\label{stringfit}
\ee
We expect $\sigma b_I^2$ to approach a finite value in the 
continuum limit ($b_I\to\infty$). 
The three parameter fit of $m(k)$ as a function of $k$
is shown in Fig.~\ref{tension}. 

\section{Extraction of $m(k)$}\label{mass}

\begin{figure} 
\centering 
%%----start of first figure---- 
\begin{minipage}[t]{0.4\linewidth} 
\centering 
\includegraphics[width=2.4in,clip=true]{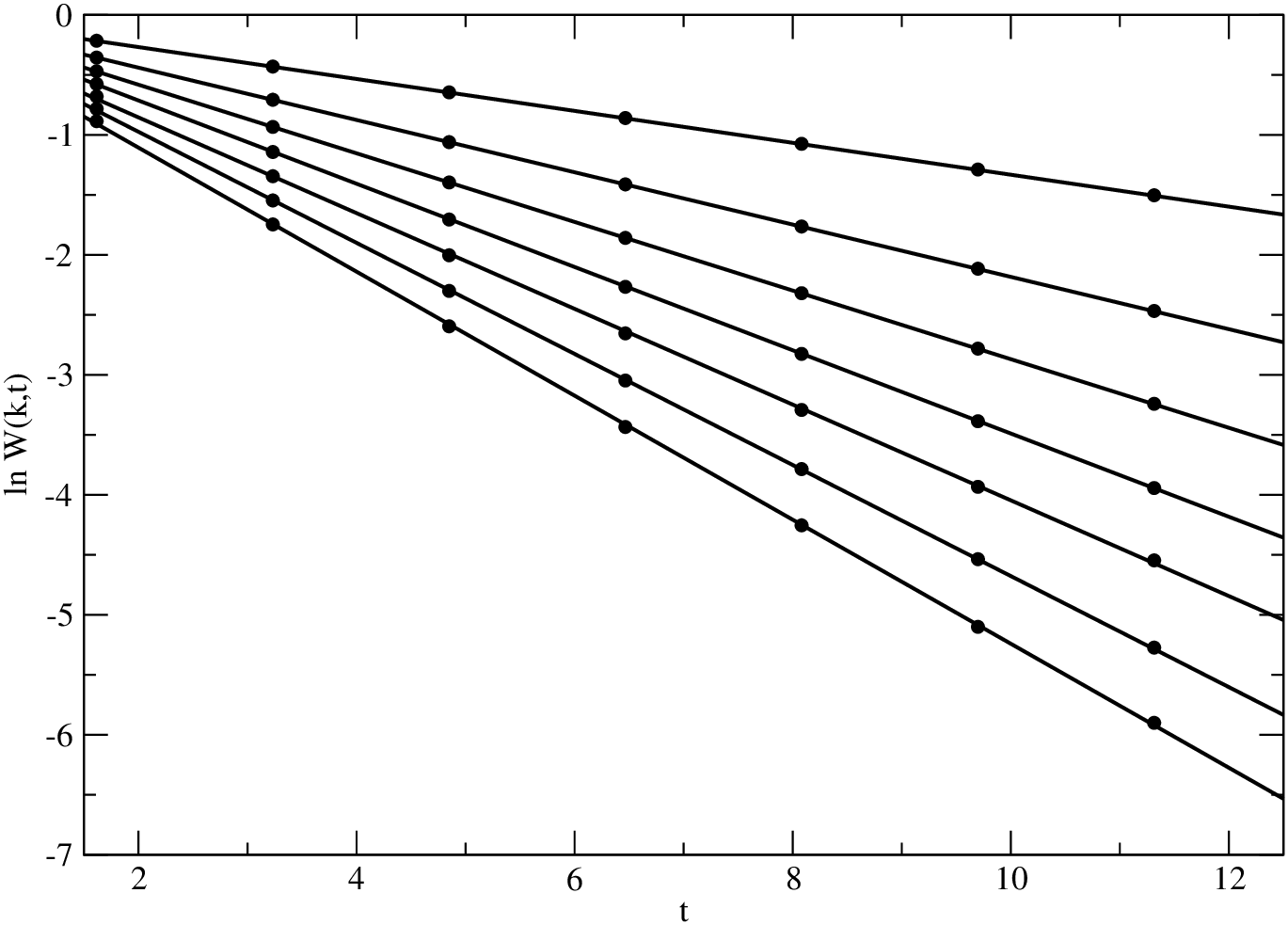} 
\caption{
Plot of $\ln W(k,t)$ as a function of $t$ for seven
different values of $k$ at $b=0.8$ with $\tau=2.5$.
}
\label{wloop1} 
\end{minipage}% 
\hspace{1cm}% 
%%----start of second figure---- 
\begin{minipage}[t]{0.4\linewidth} 
\centering 
\includegraphics[width=2.4in,clip=true]{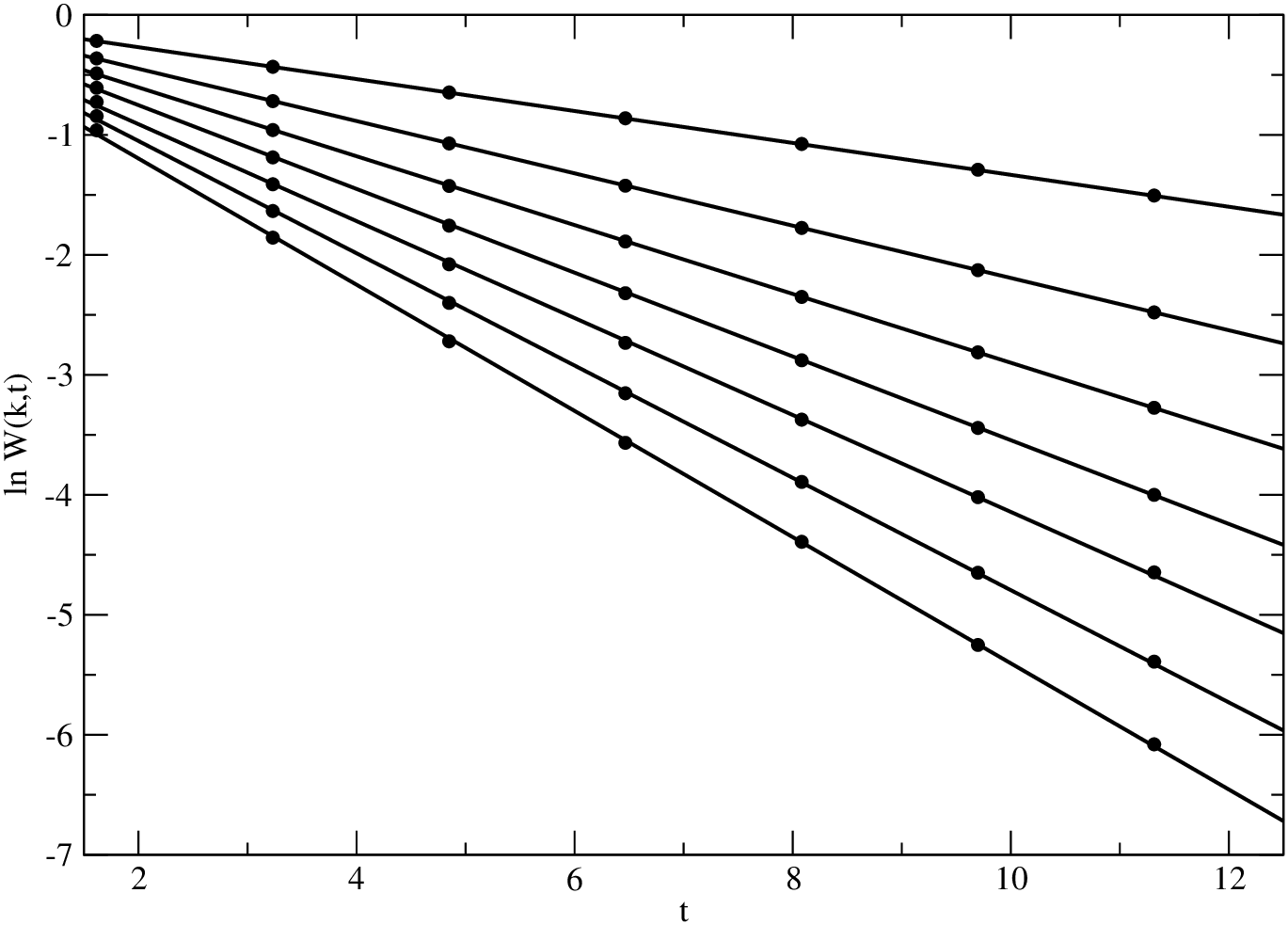} 
\caption{
Plot of $\ln W(k,t)$ as a function of $t$ for seven
different values of $k$ at $b=0.8$ with $\tau=1.25$.
} 
\label{wloop2} 
\end{minipage} 
\end{figure} 

The dimensionless ground state energy $m(k)$ is extracted
at a fixed $k$ by fitting $\ln W(k,t)$ to $-a-m(k)t$ as
discussed in Sec.~\ref{details}.
While $m(k)$ should be independent of the smearing parameter 
$\tau=fn$,
the value of $a$ 
is expected to depend on $\tau$. 

We will use $b=0.8$ as the coupling to illustrate the extraction
of $m(k)$.
Figure~\ref{wloop1} and Fig.~\ref{wloop2} show the performance
of the fit for two different values of $\tau$, namely, $2.5$
and $1.25$ respectively. The {\sl solid circles} show the data
points without errors. The {\sl solid lines} show the fit of
the data. Seven values of $t$ were used to fit the data at one
$k$, and data at seven different values of $k$ were fitted.
This amounted to all Wilson loops from $1\times 1$ to $7\times 7$
on the $5^3$ lattice. 
%The set of thermalized configurations
%used at $\tau=2.5$ is statistically independent from the
%set used at $\tau=1.25$. 
The fit parameters are shown in
Table~\ref{tab1} and Table~\ref{tab2}. Only the average
values of the fit parameters are listed. 

Investigation of Table~\ref{tab1} and Table~\ref{tab2}
shows that $m(k)$ does not depend on $\tau$. There is
a small difference in the two values of $m(k)$ at a fixed
$k$ for the two different values of $\tau$ if $k$ is
large. 
Additional analysis
shows that this difference
is within errors. Furthermore, the fitted values of
$\sigma b_I^2$ for the two different values of $\tau$
are the same within errors.

{
\TABULAR
{cccccccc}
{$k$ & 1.62 & 3.23 & 4.85 & 6.47 & 8.08 & 9.70 & 11.31 \cr
$a$ & 0.001 & 0.003 & 0.009 & 0.019 & 0.055 & 0.047 & 0.071 \cr
$m(k)$ & 0.133 & 0.218 & 0.286 & 0.347 & 0.399 & 0.464 & 0.517 \cr
}
{Fit parameters corresponding to the fit $\ln W(k,t) = -a -m(k) t$
for seven different values of $k$ at $b=0.8$ with $\tau=2.5$.
\label{tab1}}}

{
\TABULAR
{cccccccc}
{$k$ & 1.62 & 3.23 & 4.85 & 6.47 & 8.08 & 9.70 & 11.31 \cr
$a$ & 0.002 & 0.012 & 0.029 & 0.054 & 0.102 & 0.114 & 0.144 \cr
$m(k)$ & 0.133 & 0.218 & 0.287 & 0.349 & 0.404 & 0.468 & 0.526 \cr
}
{Fit parameters corresponding to the fit $\ln W(k,t) = -a -m(k) t$
for seven different values of $k$ at $b=0.8$ with $\tau=1.25$.
\label{tab2}}}

The values of $a$ in Table~\ref{tab1} and Table~\ref{tab2}
do show a variation with $\tau$ and $k$. Since a smaller
value of $\tau$ implies less smearing, the overlap with
the ground state is less for smaller $\tau$, and this
results in a larger value of $a$ at smaller $\tau$.
The value of $a$ is very close to zero for small $k$
indicating excellent overlap with the ground state for
the chosen value of $\tau$. As $k$ increases, the length
of the loop increases and the perimeter divergence has
a stronger effect. This results in a larger value of
$a$ as $k$ increases at a fixed $\tau$.

\section{Finite $N$ effects}\label{finiteN}

Two issues need to be addressed with the analysis performed
so far. We have fixed our value of $N$ assuming finite $N$
effects are small. If $N$ is not large enough, finite $N$
effects need to be addressed. In addition, we also have
to address finite volume effects since continuum reduction
is valid only in the $N\to\infty$ limit.

\begin{figure} 
\centering 
%%----start of first figure---- 
\begin{minipage}[t]{0.4\linewidth} 
\centering 
\includegraphics[width=2.4in,clip=true]{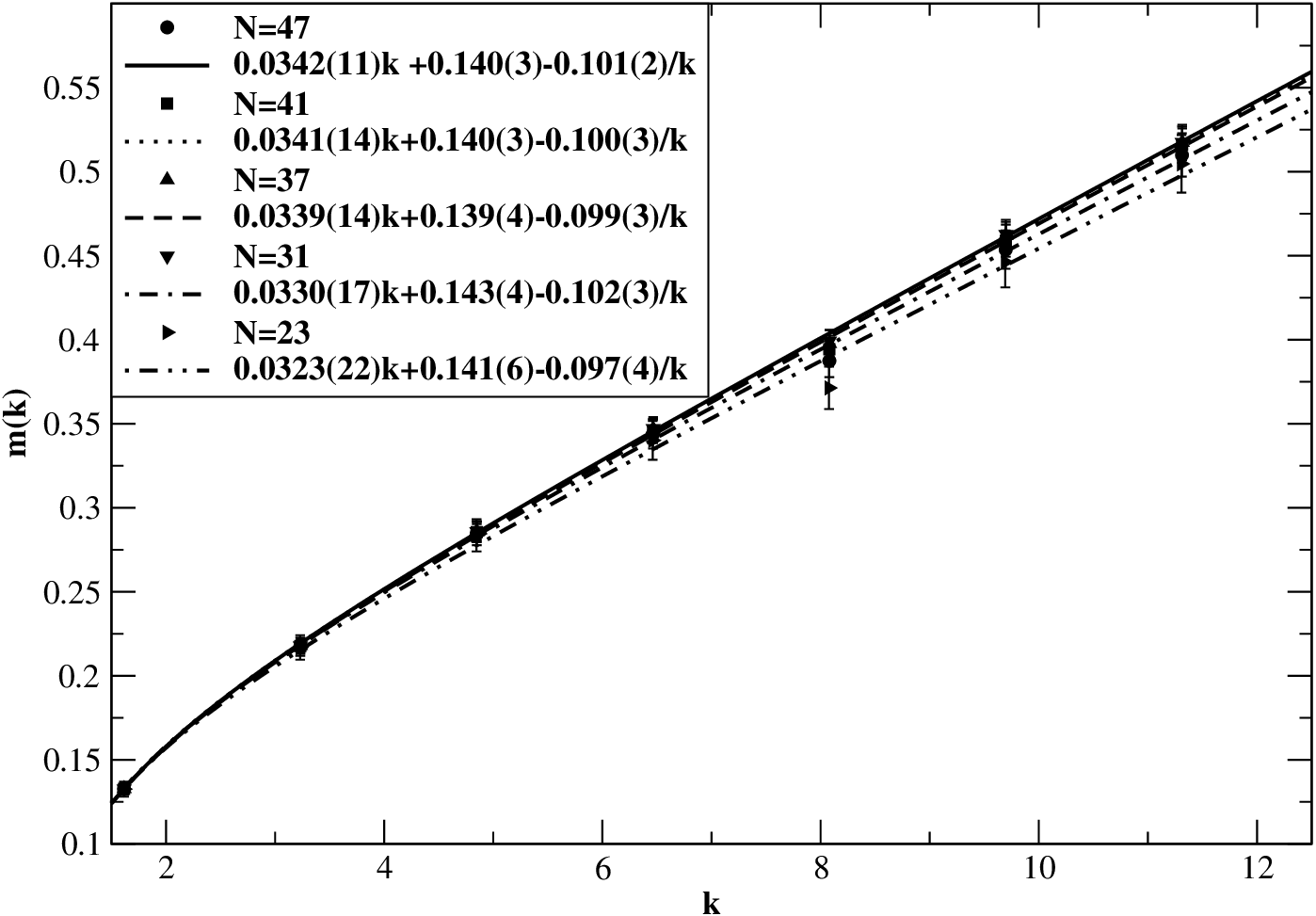} 
\caption{
The ground state energy $m(k)$ as a 
function of $k$ for five different values of $N$
}
\label{massn} 
\end{minipage}% 
\hspace{1cm}% 
%%----start of second figure---- 
\begin{minipage}[t]{0.4\linewidth} 
\centering 
\includegraphics[width=2.4in,clip=true]{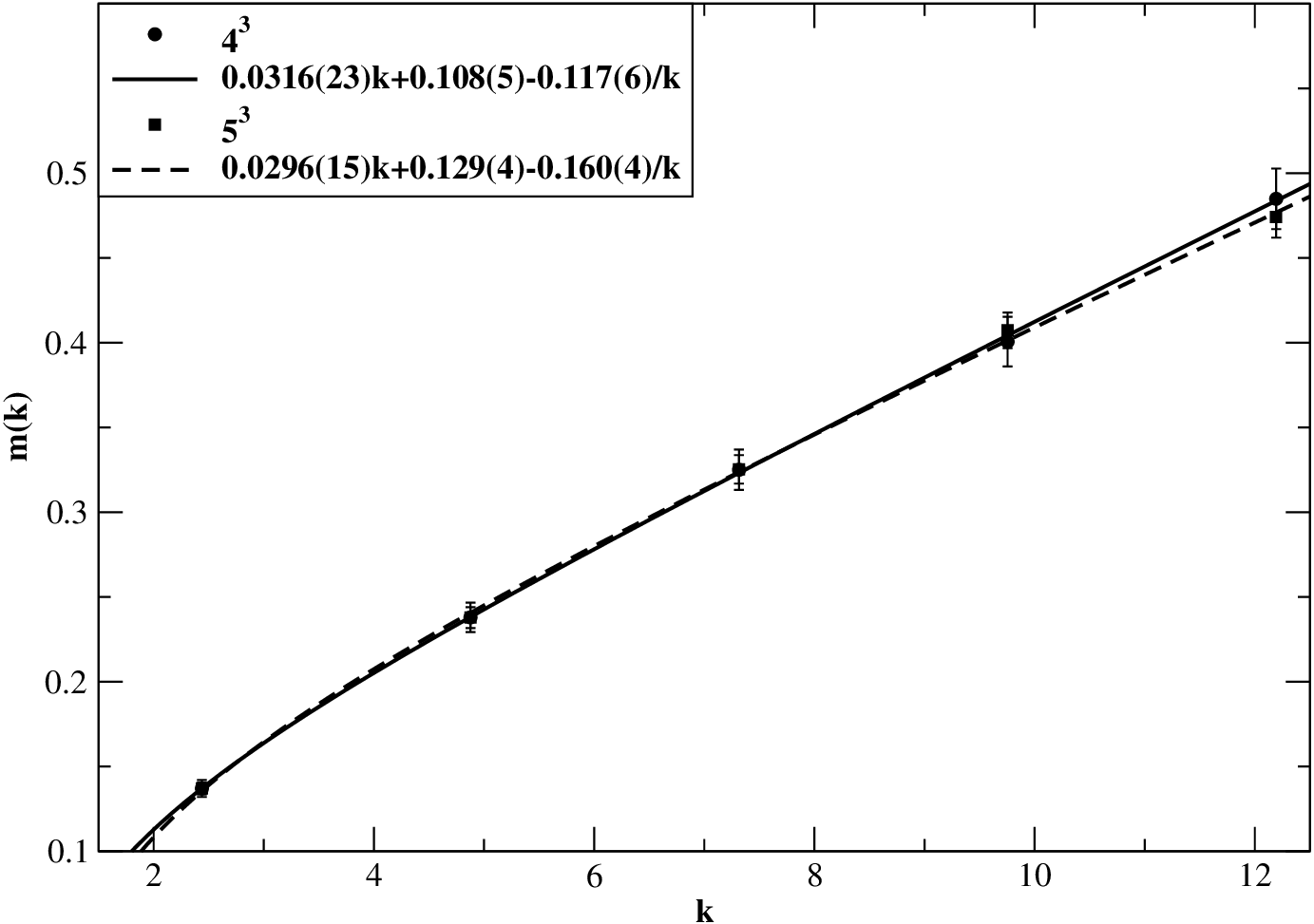} 
\caption{
The ground state energy $m(k)$ as a 
function of $k$ on two different lattices 
at $b=0.8$.
} 
\label{massl} 
\end{minipage} 
\end{figure}

We expect $m(k)$ to have a fixed limit as $N\to\infty$
at a fixed $k$, $L$, $b$ and $\tau$. Indeed, this is
the case as shown in Fig.~\ref{massn} where the results
for $m(k)$ as a function of $k$ are shown for $b=0.8$
with $\tau=2.5$ on $5^3$ lattice. All three fit parameters
are consistent within errors all the way from $N=23$ to
$N=47$. The only glitch one sees is at $k\approx 8$.
This corresponds to $K=kb_I=5$, which is
the linear extent of the lattice. One can argue that there are
larger finite $N$ effects at strong coupling
for $K=L$.
Since the fit of $m(k)$ involves several values of $k$,
the larger effect at this particular value of $k$ is
diminished in the extraction of $\sigma b_I^2$.

Since finite $N$ effects can be ignored at $N=47$, we also
expect there to be no appreciable finite volume effects at
this value of $N$. This point is illustrated in Fig.~\ref{massl}
where the result for $m(k)$ is plotted at $b=0.6$ and $\tau=2.5$
on $4^3$ and $5^3$ lattice. We used $b=0.6$ for this comparison
since we have to be in the confined phase on $4^3$ lattice.
Figure~\ref{massl} shows that the two values of $m(k)$ 
at a fixed $k$ are consistent with each other within errors.
The same is the case for the fit parameter $\sigma b_I^2$.
This is not the case for $c_1$ and $c_0 b_I$, and this is
probably due to a three parameter fit using only five data points.
Sub-leading coefficients are expected to depend sensitively on
the data points. Since we are primarily concerned with the
value of the string tension in this paper and since all our
results are based on data taken on $5^3$, we expect the
final result to be free of finite $N$ and finite $L$ errors.

\section{Creutz ratio}\label{crratio}

It is natural to ask how the Creutz ratio~\cite{Creutz:1984mg},
\be
\chi(K,J) = -\ln \frac{W(K,J)W(K-1,J-1)}{W(K,J-1)W(K-1,J)},
\ee
performs as an observable from which to extract the string tension. If 
we were to use Creutz ratios, we would have smeared all links using all 
staples. But one can still ask how the Creutz ratio behaves with the 
asymmetrically smeared links. The $K \times K$ square Creutz ratios do 
not converge well as $K$ increases. It is possible the situation would 
be different if we had smeared all links.

Each data point in 
a Creutz ratio
is obtained using only
four different Wilson loops, {\em i.e.} four of the data points in 
Fig.~\ref{wloop1}.
This is quite different from
the analysis in this paper. Seven different Wilson loops
in Fig.~\ref{wloop1} are used to extract one $m(k)$ point in 
Fig.~\ref{tension}, and the loops used
for different $k$ form independent sets. Then the $m(k)$ are fit to 
determine the string tension. Both folded and unfolded loops contribute 
together. This is the main
reason we succeeded in extracting the string tension using
the range of Wilson loops considered here. To extract the string tension 
using Creutz ratios, larger loops
and therefore larger statistics and possibly larger $N$
would be needed.

\acknowledgments

R.N. acknowledges partial support by the NSF under grant number
PHY-055375.

\end{document}